\documentclass[9pt,technote]{IEEEtran}
\usepackage{graphicx} 
\usepackage{cite,array}
\usepackage{amssymb} 
\usepackage[cmex10]{amsmath}
\newtheorem{definition}{Definition}
\newtheorem{lemma}{Lemma}
\newtheorem{theorem}{Theorem}
\newtheorem{remark}{Remark}

\newtheorem{example}{Example}

\newcommand{\set}[1]{\left\{#1\right\}}

\newcommand{\norm}[1]{\left\lVert#1\right\rVert}
\DeclareMathOperator{\E}{E}

\begin{document}

\title{Stability Analysis of Continuous-Time Switched Systems with a Random
  Switching Signal\footnote{The work in this paper was financially supported
    by National Natural Science Foundation of China under Grant 61004044,
    Program for New Century Excellent Talents in University 11-0880,
    Fundamental Research Funds for the Central Universities WK2100100013, GRF
    HKU 7138/10E, and SZSTI Basic Research Program under grant code
    JCYJ20120831142942514, and the Royal Society of Edinburgh.}}

\author{%
  Junlin Xiong,\thanks{Department of Automation, University of Science and
    Technology of China, Hefei 230026, China. \texttt{\small
      junlin.xiong@gmail.com}}  %
  James Lam,\thanks{Department of Mechanical Engineering, University of Hong
    Kong, Pokfulam Road, Hong Kong. \texttt{\small james.lam@hku.hk}}  %
  Zhan Shu,\thanks{School of Engineering Sciences, University of Southampton,
    Southampton SO17 1BJ, UK. \texttt{\small hustd8@gmail.com}}  %
  Xuerong Mao\thanks{Department of Mathematics and Statistics,
    University of Strathclyde, Glasgow G1 1XH, UK. \texttt{\small
      x.mao@strath.ac.uk}} }


\maketitle

\begin{abstract}
  This paper is concerned with the stability analysis of continuous-time
  switched systems with a random switching signal.  The switching signal
  manifests its characteristics with that the dwell time in each subsystem
  consists of a fixed part and a random part.  The stochastic stability of
  such switched systems is studied using a Lyapunov approach.  A necessary and
  sufficient condition is established in terms of linear matrix inequalities.
  The effect of the random switching signal on system stability is illustrated
  by a numerical example and the results coincide with our intuition.
\end{abstract}

\begin{IEEEkeywords}
  Dwell time, random switching, stochastic stability.
\end{IEEEkeywords}

\section{Introduction}
\label{sec:introduction}

Generally speaking, a switched system is a dynamical system that consists of a
finite number of subsystems and a switching signal. The subsystems are
described by differential equations and are employed to capture the dominant
dynamics of the system in different operation modes. The switching signal
decides which subsystem is being activated (equivalently, which operation mode
the system is working in) at a particular time.  The study on switched systems
has attracted a lot of research attention
\cite{WLWC:2011:laji,SW:2012,JTLD:2012,ZHGW:2012}.  Switched systems have
various applications such as in communication networks \cite{Hes:2005:na},
aerospace industry \cite{Boukas:2006:book} and networked control systems
\cite{YSLG:2011:auto,LZBT:2012:jfi,ZGK:2013:tii}. The readers are referred to
\cite{Liberzon:2003:book,Mao:2006:book} for a general introduction and
\cite{LA:2009:tac} for a recent review.

A special class of switched systems with a random switching signal is
Markovian jump systems where the switching signal is modeled by a Markov
process \cite{Mao:2006:book}. The sliding mode control of Markovian jump
systems has been studied in \cite{WSS:2012:auto}. The filtering problem has
been investigated in \cite{WGW:2008:auto}.  Some results of Markovian jump
systems with time delays have been reported in \cite{SBA:1999:taca}.  When
there are switching probability uncertainties, the stochastic stability
problems have been studied in \cite{XLG+:2005:auto,ZL:2010:tac,ZHY+:2011:auto}
recently.

In this paper, a new class of random switching signals is proposed to activate
the subsystems of switched systems, and a necessary and sufficient condition
is established for the stochastic stability analysis. For switched systems
with the switching signal proposed in this paper, the dwell time in each
subsystem consists of two parts: the fixed dwell time and the random dwell
time. The fixed dwell time plays a similar role as the ``dwell time'' in
deterministic switched systems~\cite{Morse:1996:tac}; the random dwell time is
corresponding to the exponentially distributed ``sojourn time'' in Markovian
jump systems~\cite{JC:1990:tac}. With the proposed class of random switching
signals, the switched system can be transformed to a Markovian jump system
with state jumps at the switching time instants.  The stochastic stability
problem is then studied using a Lyapunov approach; and a necessary and
sufficient condition is obtained. When the parameters of the random switching
signal are known, the system stability can be checked by solving a set of
coupled linear matrix inequalities. A numerical example is used to illustrate
the effect of the random switching signals on system stability. The stability
regions and instability regions are numerically determined for different
values of the fixed dwell time parameters. The numerical results demonstrate
that (1) when all the subsystems are stable, fast switching may destabilize
the system, and hence it should be avoided; (2) when both stable and unstable
subsystems are present, dwelling in the stable subsystems longer can increase
the degree of the stability, otherwise the system will tend to become
unstable; (3) when all the subsystems are unstable, both fast and slow
switching can destabilize the system, the system stability, however, may
sometimes be achieved by choosing the fixed dwell time parameters properly.

Compared to the previous work in
\cite{JC:1990:tac,SBA:1999:taca,XLG+:2005:auto,Boukas:2006:book,Mao:2006:book,WGW:2008:auto,ZL:2010:tac,ZHY+:2011:auto,WSS:2012:auto},
the work in this paper provides a new and more general view of switching, and
the corresponding stability results. The class of random switching signals in
this paper allows that a fixed dwell time can exist for each mode before a
Markov switch occurs. Hence, the systems in this paper can possibly accommodate more
realistic situations; the results in this paper should be applicable, in
principle, to all previous cases. Moreover, the results in this paper also lay
a foundation for novel hybrid controller design, which is illustrated by
the numerical example in Section~\ref{sec:illustr-exampl}.

\textit{Notation:} $\mathbb{R}^n$ and $\mathbb{S}^{+}$ are, respectively, the
$n$-dimensional Euclidean space and the set of $n\times n$ real symmetric
positive definite matrices. Notation $X<Y$, where $X$ and $Y$ are real
symmetric matrices, means that the matrix $X-Y$ is negative definite. The
superscript ``$T$'' denotes the transpose for vectors or
matrices. $\norm{\cdot}$ refers to the Euclidean norm for
vectors. $\overline{\sigma}(\cdot)$ and $\underline{\sigma}(\cdot)$ are,
respectively, the maximum and the minimum singular values of square
matrices. Moreover, let $(\Omega, \mathcal{F}, \mathrm{Pr})$ be a complete
probability space. $\E\{\cdot\}$ and $\sigma\{\cdot\}$ stand for the
expectation and the generated $\sigma$-algebra, respectively.

\section{Problem Formulation}
\label{sec:problem-formulation}

Consider a class of switched linear systems defined on a complete probability
space $(\Omega, \mathcal{F}, \mathrm{Pr})$
\begin{equation}
  \label{eq:1}
  \dot{x}(t) = A_{r(t)}x(t)
\end{equation}
where $x(t)\in\mathbb{R}^{n}$, $t\ge0$, is the system state, $r(t) \in
\mathcal{M} \triangleq \set{1,2,\ldots,m}$ is the switching signal deciding
the current system operation mode.  Suppose the system switches its operation
mode to $i\in\mathcal{M}$ at time $t_{k}$, the characteristic of the switching
signal $r(t)$ can be described as follows. For time $t \in
[t_{k},t_{k}+d_{i})$, where $d_{i} \ge 0$, no switching is allowed almost
surely; that is,
\begin{equation}
  \label{eq:2}
  \Pr
  \left\{
    r(t+\Delta t)=j \mid r(t)=i
  \right\}
  =
  \left\{
    \begin{aligned}
      &0, \text{ if $j\ne i$} \\
      &1, \text{ if $j=i$}
    \end{aligned}
  \right.
\end{equation}
where $\Delta t$ is a small time increment satisfying $\lim_{\Delta
  t\to0^{+}}\frac{o(\Delta t)}{\Delta t}=0$.  The parameter $d_{i}$ plays the
role of ``dwell time'' in deterministic switched systems\cite{Morse:1996:tac},
and is called the fixed dwell time of the system in \eqref{eq:1}.  For $t \ge
t_{k}+d_{i}$, mode switching occurs according to the mode transition
probabilities given by
\begin{equation}
  \label{eq:3}
  \Pr
  \left\{
    r(t+\Delta t)=j \mid r(t)=i
  \right\}
  =
  \left\{
    \begin{aligned}
      &\pi_{ij}\Delta t+o(\Delta t), \text{if $j\ne i$} \\
      &1+\pi_{ii}\Delta t+o(\Delta t), \text{if $j=i$}
    \end{aligned}
  \right.
\end{equation}
where $\pi_{ij}\ge0$ if $j\ne i$, and $\pi_{ii}\triangleq-\sum_{j=1,j\ne
  i}^{m}\pi_{ij}$ if $j=i$. If the next switching occurs at time $t_{k+1}$, we can
define $\eta_{i} \triangleq t_{k+1}-(t_{k}+d_{i})$, which is an exponential
random variable with parameter $v_{i} \triangleq -\pi_{ii}$ according to
\eqref{eq:3}. To simplify the derivation of the main results, the system
\eqref{eq:1} is assumed to have no absorbing mode; that is, $\pi_{ii} \ne 0$
for all $i\in\mathcal{M}$. The random variable $\eta_{i}$ plays the role of
``sojourn time'' in Markovian jump systems\cite{Mao:2006:book,JC:1990:tac},
and is called the random dwell time of the system.  The dwell time of
system~\eqref{eq:1} in mode $i$ is defined as $\tau_{i} \triangleq
t_{k+1}-t_{k}=d_{i}+\eta_{i}$, indicating the total time length of the system
\eqref{eq:1} being in mode $i$. As a result, the time interval
$[t_{k},t_{k+1})$ can be correspondingly divided into two parts:
$[t_{k},t_{k+1}) = [t_{k},t_{k}+d_{i}) \cup [t_{k}+d_{i},t_{k+1})$. It can be
seen from above that $t_{0}=0$, $t_{k+1}>t_{k}$ and
$\lim_{k\to\infty}t_{k}=\infty$.

\begin{example}
  Let us illustrate the property of the switching signal $r(t)$ with an
  example.  Suppose $r(t)\in\set{1,2,3}$. A sample path of $r(t)$ is
  illustrated in Fig.~\ref{fig:dwellmkv}.  Here, the system changes its mode
  at time $t_{k}$ from Mode 1 to Mode 3. In view of \eqref{eq:2}, there will
  be no switching almost surely during the interval $[t_{k},t_{k}+d_3)$; the
  system is allowed to switch modes after the time $t_{k}+d_{3}$ and obeys the
  switching rule in~\eqref{eq:3}.
  \begin{figure}[bht]
    \centering
    \includegraphics[width=8cm]{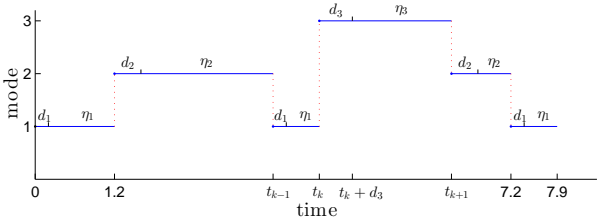}
    \caption{A sample path of the random switching signal}
    \label{fig:dwellmkv}
  \end{figure}
\end{example}

\begin{definition}
  \label{dfn:1}
  Let $x(t)$ be the state trajectory of system~\eqref{eq:1}. Then
  system~\eqref{eq:1} is said to be stochastically stable if
  \begin{equation}
    \label{eq:4}
    \E
    \left\{
      \int_{0}^{\infty} \norm{x(t)}^{2} dt
      \mid
      x_{0},r_{0}
    \right\}
    < \infty
  \end{equation}
  for any initial system state $x_{0}\in\mathbb{R}^{n}$ and any initial
  operation mode $r_{0}\in\mathcal{M}$.
\end{definition}

\begin{remark}
  The above stochastic stability definition is analogous to that of Markovian
  jump systems~\cite{JC:1990:tac}. It is also a uniform stability in the
  sense that the inequality \eqref{eq:4} is required to be true over all the
  switching signals defined by \eqref{eq:2}--\eqref{eq:3}.
\end{remark}

\section{Stability Analysis}
\label{sec:stability-analysis}

The stability property of the system in~\eqref{eq:1} is studied via two
steps. In Step 1, the stochastic stability of the system in~\eqref{eq:1} is
shown to be equivalent to the stochastic stability of an auxiliary system. In
Step 2, the stability of the auxiliary system is studied by a Lyapunov
approach, and a necessary and sufficient condition is established.

\subsection{Switched Systems with State Jumps}
\label{sec:an-equivalent-system}

A switched system with state jumps is the auxiliary system to be constructed
in this section.  The stability of the constructed auxiliary system is shown to
be equivalent to that of the system in~\eqref{eq:1}.

\begin{figure}[hbt]
  \centering
  \includegraphics[width=7cm]{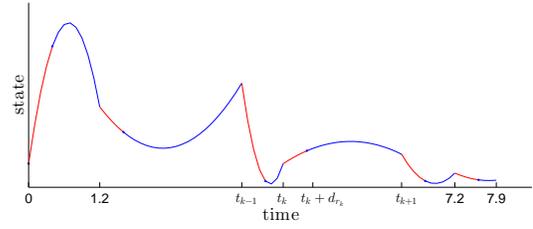}
  \caption{State trajectory of the switched systems}
  \label{fig:state}
\end{figure}

\begin{figure}[hbt]
  \centering
  \includegraphics[width=7cm]{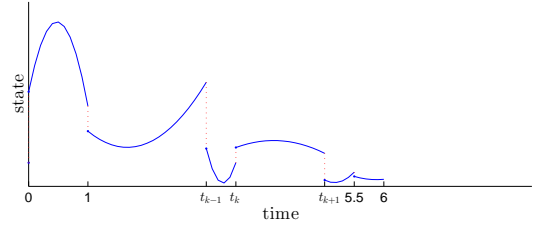}
  \caption{State trajectory of the switched systems with state jumps}
  \label{fig:jumpstate}
\end{figure}

Let us first study the state trajectory of system~\eqref{eq:1} to motivate the
construction of the auxiliary system. A sample path of the state trajectory of
system~\eqref{eq:1} is illustrated in Fig.~\ref{fig:state}. Suppose the system
switches to mode $r_{k} \in \mathcal{M}$ at time $t_{k}$. Then the system state
$x(t)$ will evolve from $x(t_{k})$ at time $t_{k}$ to $x(t_{k}+d_{r_{k}}) =
e^{A_{r_{k}}d_{r_{k}}}x(t_{k})$ at time $t_{k}+d_{r_{k}}$ almost surely.  The
idea here is to squeeze the interval $[t_{k},t_{k}+d_{r_{k}})$ to a point
$t_{k}$ and make the system in~\eqref{eq:1} having a state jump from
$x(t_{k}^{-})$ to $e^{A_{r_{k}}d_{r_{k}}}x(t_{k}^{-})$ at time $t_{k}$, as
illustrated in Fig.~\ref{fig:jumpstate}, where $x(t_{k}^{-}) \triangleq \lim_{
  t\to t_{k}^{-}} x(t)$.
As a result, the system in~\eqref{eq:1} can be transformed to a randomly
switched system with state jumps:
\begin{equation}
  \label{eq:5}
  \left\{
    \begin{aligned}
      \dot{\xi}(\tilde{t}) &= A_{\rho(\tilde{t})}\xi(\tilde{t}),
      \qquad\quad \tilde{t}_{k} < \tilde{t} < \tilde{t}_{k+1} \\
      \xi(\tilde{t}_{k}) &= e^{A_{\rho_{k}}d_{\rho_{k}}}\xi(\tilde{t}_{k}^{-}),
      \quad k = 0,1,2,\ldots
    \end{aligned}
  \right.
\end{equation}
where $\xi(\tilde{t})$ is the system state, $\tilde{t}_{k}$, $k=0,1,2,\ldots$,
are the time instants when the system switches its operation
modes. $\xi(\tilde{t}_{k}^{-}) \triangleq \lim_{t \to t_{k}^{-}}
\xi(\tilde{t})$ is the system state just before switches. $\rho(\tilde{t})$ is
a Markovian process taking values in $\mathcal{M}$ and equipped with the
transition probabilities
\begin{equation}
  \label{eq:6}
  \Pr
  \left\{
    \rho(\tilde{t}+\Delta t)=j
    \mid
    \rho(\tilde{t})=i
  \right\}
  =
  \left\{
    \begin{aligned}
      &\pi_{ij}\Delta t+o(\Delta t), \text{if $j\ne i$} \\
      &1+\pi_{ii}\Delta t+o(\Delta t), \text{if $j=i$}
    \end{aligned}
  \right.
\end{equation}
where $\pi_{ij}$ are the same as those in \eqref{eq:3}.  The mode transition
rate matrix is denoted by $\Pi \triangleq (\pi_{ij}) \in \mathbb{R}^{m\times
  m}$.  Suppose that the system~\eqref{eq:5} jumps to mode $\rho_{k}
\triangleq \rho(\tilde{t}_{k})$ at time $\tilde{t}_{k}$, the dwell time of
system~\eqref{eq:5} in mode $\rho_{k}$ is defined as $\eta_{\rho_{k}}
\triangleq \tilde{t}_{k+1}-\tilde{t}_{k}$. It follows from \eqref{eq:6} that
$\eta_{\rho_{k}}$ is an exponentially distributed random variable with
parameter $v_{\rho_{k}}$. In other words, $\eta_{\rho_{k}}$ has the same
distribution as the corresponding random dwell time in~\eqref{eq:3}.  Also, we
have that $\tilde{t}_{0}=0$, $\tilde{t}_{k+1}>\tilde{t}_{k}$,
$\lim_{k\to\infty}\tilde{t}_{k}=\infty$ and $\tilde{t}_{k+1} =
\sum_{l=0}^{k}\eta_{\rho_{l}}$.

For the systems in \eqref{eq:1} and \eqref{eq:5}, we define the filtrations
$\mathcal{F}_{t} \triangleq \sigma\set{(x(\tau),r(\tau)) : 0\le\tau\le t}$ and
$\mathcal{G}_{t} \triangleq \sigma\set{(\xi(\tau),\rho(\tau)) : 0\le\tau\le
  t}$ for $t\ge 0$, respectively. Now we are ready to establish the
equivalence of the stability properties between system~\eqref{eq:1} and
system~\eqref{eq:5}.
\begin{lemma}
  \label{lem:1}
  The stochastic stability of the system in~\eqref{eq:1} is equivalent to the
  stochastic stability of the system in~\eqref{eq:5}.
\end{lemma}

\begin{IEEEproof}
  The equivalence is proved based on the following observation: Given any
  sample path $(\xi(\tilde{t}),\rho(\tilde{t}))$ of the system
  in~\eqref{eq:5}, there is a corresponding sample path $(x(t),r(t))$ of the
  system in~\eqref{eq:1}; and vice versa. Furthermore, the two sample paths
  satisfy the following properties: For $k=0,1,2,\ldots$,
  \begin{enumerate}
  \item $x(0)=\xi(0^{-})$, $\xi(0) = e^{A_{\rho_{0}}d_{\rho_{0}}}\xi(0^{-})$
    and $t_{0}=\tilde{t}_{0}=0$.
  \item $r_{k}=r(t_{k})=\rho(\tilde{t}_{k})=\rho_{k}$.
  \item $t_{k+1} = \sum_{l=0}^{k}(d_{\rho_{l}}+\eta_{\rho_{l}}) =
    \tilde{t}_{k+1}+\sum_{l=0}^{k}d_{\rho_{l}}$.
  \item $\xi(\tilde{t}_{k}+\tau) = x(t_{k}+d_{r_{k}}+\tau)$ for $0\le\tau<\eta_{\rho_{k}}$.
  \end{enumerate}
  The fourth property is a direct result of the first three properties; that is,
  $\xi(\tilde{t}_{k}+\tau)
  = e^{A_{\rho_{k}} \tau} \xi(\tilde{t}_{k})
  = e^{A_{\rho_{k}} \tau} e^{A_{\rho_{k}} d_{\rho_{k}}} \xi(\tilde{t}_{k}^{-})
  = e^{A_{r_{k}} (d_{r_{k}} + \tau)} x(t_{k})
  = x(t_{k}+d_{r_{k}}+\tau)$.

  ($\Rightarrow$) %
  Suppose that the system in~\eqref{eq:1} is stochastically stable, the
  stability of the system in~\eqref{eq:5} follows from
  \begin{align*}
    &\quad \E
    \left\{
      \int_{0}^{\infty} \norm{\xi(\tilde{t})}^{2} d\tilde{t}
      \mid
      \xi_{0}, \rho_{0}
    \right\} \\
    &= \E
    \left\{
      \sum_{k=0}^{\infty}
      \left\{
        \int_{\tilde{t}_{k}}^{\tilde{t}_{k+1}}
        \norm{\xi(\tilde{t})}^{2} d\tilde{t}
      \right\}
      \mid
      \xi_{0}, \rho_{0}
    \right\} \\
    &= \E
    \left\{
      \sum_{k=0}^{\infty}
      \left\{
        \int_{t_{k}+d_{r_{k}}}^{t_{k+1}} \norm{x(t)}^{2} dt
      \right\}
      \mid x_{0}, r_{0}
    \right\} \\
    &\le \E
    \left\{
      \sum_{k=0}^{\infty}
      \left\{
        \int_{t_{k}}^{t_{k+1}} \norm{x(t)}^{2} dt
      \right\}
      \mid x_{0}, r_{0}
    \right\} \\
    &= \E
    \left\{
      \int_{0}^{\infty}\norm{x(t)}^{2} dt \mid x_{0}, r_{0}
    \right\} < \infty
  \end{align*}
  The second ``$=$'' holds because of the fourth property of the two
  sample paths.

  ($\Leftarrow$) %
  Suppose that the system in~\eqref{eq:5} is stochastically stable.  Note that
  $\eta_{r_{k}}$ is an exponential random variable with parameter $v_{r_{k}}$.
  In view of Lemma~\ref{lem:2} and Lemma~\ref{lem:3} (see Appendix), there
  exists a real number $c_{r_{k}}<v_{r_{k}}$ such that
  \begin{align}
    &\quad \E\left\{
      \int_{t_{k}+d_{r_{k}}}^{t_{k+1}} \norm{x(t)}^{2}dt
      \mid \mathcal{F}_{t_{k}}
    \right\} \notag \\
    &= \E\left\{
      \int_{t_{k}+d_{r_{k}}}^{t_{k+1}} \norm{x(t)}^{2}dt
      \mid x(t_{k}), r(t_{k})=r_{k}
    \right\} \notag \\
    &= \E\left\{
      \int_{t_{k}+d_{r_{k}}}^{t_{k}+d_{r_{k}}+\eta_{r_{k}}}
      \norm{e^{A_{r_{k}}(t-t_{k})}x(t_{k})}^{2}dt
    \right\} \notag  \\
    &= \E\left\{
      \int_{d_{r_{k}}}^{d_{r_{k}}+\eta_{r_{k}}}
      \norm{e^{A_{r_{k}}\tau}x(t_{k})}^{2}d\tau
    \right\} \notag \\
    &\ge \E\left\{
      \int_{d_{r_{k}}}^{d_{r_{k}}+\eta_{r_{k}}}
      e^{c_{r_{k}}\tau}\norm{x(t_{k})}^{2}d\tau
    \right\} \notag \\
    &= \E\left\{
      \int_{d_{r_{k}}}^{d_{r_{k}}+\eta_{r_{k}}}
      e^{c_{r_{k}}\tau}d\tau
    \right\} \norm{x(t_{k})}^{2} \notag  \\
    &= \frac{e^{v_{r_{k}}d_{r_{k}}}}{v_{r_{k}}-c_{r_{k}}}\norm{x(t_{k})}^{2} 
    \ge \alpha_{\mathrm{min}}\norm{x(t_{k})}^{2}
    \label{eq:7}
  \end{align} 
  for any $x(t_{k})\ne0$ and any $r_{k}\in\mathcal{M}$, where $
  \alpha_{\mathrm{min}} \triangleq \min_{i\in\mathcal{M}} \set{
    \frac{e^{v_{i}d_{i}}}{v_{i}-c_{i}} }>0 $. The first ``$=$'' holds due to
  the property of the process $(x(t),r(t))$, the first ``$\ge$'' holds because
  of Lemma~\ref{lem:2}, and the last ``$=$'' holds because of
  Lemma~\ref{lem:3}.  Now, we have
  \begin{align}
    &\quad \E
    \left\{
      \int_{t_{k}}^{t_{k}+d_{r_{k}}}\norm{x(t)}^{2}dt
      \mid \mathcal{F}_{t_{k}}
    \right\} \notag \\
    &= \E
    \left\{
      \int_{t_{k}}^{t_{k}+d_{r_{k}}}\norm{x(t)}^{2}dt
      \mid x(t_{k}), r(t_{k})=r_{k}
    \right\} \notag \\
    &= \int_{t_{k}}^{t_{k}+d_{r_{k}}} \norm{e^{A_{r_{k}}(t-t_{k})}x(t_{k})}^{2}dt \notag \\
    &= \int_{0}^{d_{r_{k}}}\norm{e^{A_{r_{k}}\tau}x(t_{k})}^{2} d\tau \notag \\
    &\le \int_{0}^{d_{r_{k}}} \alpha_{\mathrm{max}}\norm{x(t_{k})}^{2} d\tau \notag  \\
    &\le d_{\mathrm{max}}\alpha_{\mathrm{max}}\norm{x(t_{k})}^{2} \notag  \\
    &\le \frac{d_{\mathrm{max}}\alpha_{\mathrm{max}}}{\alpha_{\mathrm{min}}}
    \E
    \left\{
      \int_{t_{k}+d_{r_{k}}}^{t_{k+1}} \norm{x(t)}^{2}dt
      \mid \mathcal{F}_{t_{k}}
    \right\}
    \label{eq:8}
  \end{align}
  where $ \alpha_{\mathrm{max}} \triangleq \max_{i\in\mathcal{M}}\set{
    \max_{\tau\in[0,d_{i}]} \overline{\sigma}^{2}\left(e^{A_{i}\tau}\right)
  }$ and $ d_{\mathrm{max}} \triangleq \max_{i\in\mathcal{M}} \{d_{i}\} $.  The
  last ``$\le$'' holds because of \eqref{eq:7}.  Therefore,
  \begin{align}
    &\quad \E\left\{
      \int_{t_{k}}^{t_{k+1}}\norm{x(t)}^{2}dt \mid
      \mathcal{F}_{t_k}
    \right\} \notag \\
    &= \E\left\{
      \int_{t_{k}}^{t_{k}+d_{r_{k}}}\norm{x(t)}^{2}dt \mid
      \mathcal{F}_{t_k}
    \right\}
    + \E\left\{
      \int_{t_{k}+d_{r_{k}}}^{t_{k+1}}\norm{x(t)}^{2}dt \mid
      \mathcal{F}_{t_k}
    \right\} \notag \\
    &\le \left[\frac{d_{\mathrm{max}}\alpha_{\mathrm{max}}}{\alpha_{\mathrm{min}}}+1\right]
    \E\left\{
      \int_{t_{k}+d_{r_{k}}}^{t_{k+1}}\norm{x(t)}^{2}dt
      \mid \mathcal{F}_{t_k}
    \right\} \notag \\
    &= \left[\frac{d_{\mathrm{max}}\alpha_{\mathrm{max}}}{\alpha_{\mathrm{min}}}+1\right]
    \E\left\{
      \int_{\tilde{t}_{k}}^{\tilde{t}_{k+1}}\norm{\xi(\tilde{t})}^{2}d\tilde{t}
      \mid \mathcal{G}_{\tilde{t}_{k}} 
    \right\}
    \label{eq:9}
  \end{align}
  Here, the ``$\le$'' holds because of \eqref{eq:8}, and the last ``$=$''
  holds because of the four properties stated in the beginning of the proof.

  But, by the property of the conditional expectation, we have 
  \begin{align*}
    &\E\left\{
      \E\left\{
        \int_{t_{k}}^{t_{k+1}} \norm{x(t)}^{2} dt
        \mid \mathcal{F}_{t_k}
      \right\}
      \mid x_{0}, r_{0}
    \right\} \\
    &\qquad \qquad = \E\left\{
      \int_{t_{k}}^{t_{k+1}}\norm{x(t)}^{2}dt\mid x_{0}, r_{0}
    \right\}
  \end{align*}
  and
  \begin{align*}
    & \E\left\{
      \E\left\{
        \int_{\tilde{t}_{k}}^{\tilde{t}_{k+1}}
        \norm{\xi(\tilde{t})}^{2}d\tilde{t} \mid \mathcal{G}_{\tilde{t}_{k}}
      \right\}
      \mid \xi_{0}, \rho_{0}
    \right\} \\
    &\qquad\qquad= \E\left\{
      \int_{\tilde{t}_{k}}^{\tilde{t}_{k+1}}
      \norm{\xi(\tilde{t})}^{2}d\tilde{t} \mid \xi_{0}, \rho_{0}
    \right\}
  \end{align*}
  Taking the conditional expectation on both sides of \eqref{eq:9}, we have
  \begin{align*}
    & \E
    \left\{
      \int_{t_{k}}^{t_{k+1}}\norm{x(t)}^{2}dt \mid x_{0}, r_{0}
    \right\} \\
    &\qquad\qquad\le
    \left[
      \frac{d_{\mathrm{max}}\alpha_{\mathrm{max}}}{\alpha_{\mathrm{min}}}+1
    \right]
    \E
    \left\{
      \int_{\tilde{t}_{k}}^{\tilde{t}_{k+1}}\norm{\xi(\tilde{t})}^{2}d\tilde{t}
      \mid \xi_{0}, \rho_{0}
    \right\}
  \end{align*}

  Finally, the stability of the system in~\eqref{eq:5} implies that
  \begin{align*}
    &\quad \E
    \left\{
      \int_{0}^{\infty} \norm{x(t)}^{2} dt \mid x_{0}, r_{0}
    \right\} \\
    &= \E
    \left\{
      \sum_{k=0}^{\infty}\int_{t_{k}}^{t_{k+1}} \norm{x(t)}^{2} dt \mid x_{0},
      r_{0}
    \right\} \\
    &= \sum_{k=0}^{\infty}\E
    \left\{
      \int_{t_{k}}^{t_{k+1}} \norm{x(t)}^{2} dt \mid x_{0}, r_{0}
    \right\} \\
    &\le
    \left[
      \frac{d_{\mathrm{max}}\alpha_{\mathrm{max}}}{\alpha_{\mathrm{min}}}+1
    \right]
    \sum_{k=0}^{\infty} \E
    \left\{
      \int_{\tilde{t}_{k}}^{\tilde{t}_{k+1}}\norm{\xi(\tilde{t})}^{2}d\tilde{t}
      \mid \xi_{0}, \rho_{0}
    \right\} \\
    &=
    \left[
      \frac{d_{\mathrm{max}}\alpha_{\mathrm{max}}}{\alpha_{\mathrm{min}}}+1
    \right]
    \E
    \left\{
      \int_{0}^{\infty}\norm{\xi(\tilde{t})}^{2}d\tilde{t} \mid \xi_{0},
      \rho_{0}
    \right\} \\
    &< \infty
  \end{align*}
  Therefore, the system in~\eqref{eq:1} is stochastically stable. 
\end{IEEEproof}

\begin{remark}
  Lemma~\ref{lem:1} still holds if the assumption of no absorbing mode
  existing in the system is removed. In this case, the proof is still valid
  before the systems enter a absorbing mode. Once the systems enter a
  absorbing mode, the subsystem corresponding to the absorbing mode must be a
  stable subsystem.
\end{remark}

In view of Lemma~\ref{lem:1}, the study of stability of the system
in~\eqref{eq:1} is transferred to the stability study of
system~\eqref{eq:5}. Since the switching signal of system~\eqref{eq:5} is a
Markov process, the stability analysis becomes solvable.

\subsection{Stability Result}
\label{sec:stability-result}

In this section, a necessary and sufficient condition is derived for the
stochastic stability analysis of system~\eqref{eq:1} based upon
Lemma~\ref{lem:1}.

\begin{theorem}
  \label{thm:1}
  System~\eqref{eq:1} is stochastically stable if and only if there exist
  matrices $P_{i}\in\mathbb{S}^{+}$, $i\in\mathcal{M}$, such that
  \begin{equation}
    \label{eq:10}
    A_{i}^{T}P_{i}+P_{i}A_{i}+\pi_{ii}P_{i}
    +\sum_{j=1,j\ne i}^{m}\left\{\pi_{ij}e^{A_{j}^{T}d_{j}}P_{j}e^{A_{j}d_{j}}\right\}
    < 0
  \end{equation}
  for all $i\in\mathcal{M}$.
\end{theorem}

\begin{IEEEproof}
  ($\Leftarrow$) 
  Suppose that there exist matrices $P_{i}\in\mathbb{S}^{+}$ such that
  \eqref{eq:10} holds for all $i\in\mathcal{M}$, we shall show that the system
  in~\eqref{eq:5} is stochastically stable. Consider the Lyapunov function
  \begin{align*}
    V(\xi(\tilde{t}),\rho(\tilde{t})) \triangleq
    \xi^{T}(\tilde{t})P(\rho(\tilde{t}))\xi(\tilde{t})
  \end{align*}
  where $P(\rho(\tilde{t})=i) \triangleq P_{i}>0$.  It follows from \eqref{eq:5} and
  \eqref{eq:6} that
  \begin{align}
    &\Pr\left\{
      \xi(\tilde{t}+\Delta t)
      =e^{A_{j}d_{j}}[\xi(\tilde{t})+A_{j}\xi(\tilde{t})\Delta t] + o(\Delta t)
      \mid \xi(\tilde{t}), \rho(\tilde{t})=i
    \right\} \notag \\
    &\qquad\qquad\qquad\qquad\qquad = \pi_{ij}\Delta t+o(\Delta t), \text{
      if $j\ne i$}
    \label{eq:11} \\
    &\Pr\left\{
      \xi(\tilde{t}+\Delta t)=\xi(\tilde{t})+A_{i}\xi(\tilde{t})\Delta
      t+o(\Delta t)
      \mid \xi(\tilde{t}), \rho(\tilde{t})=i
    \right\} \notag \\
    &\qquad\qquad\qquad\qquad\qquad =
    1+\pi_{ii}\Delta t+o(\Delta t), \text{ if $j=i$}
    \label{eq:12} \\
    &\Pr\left\{
      P(\rho(\tilde{t}+\Delta t))=P_{j} \mid \xi(\tilde{t}),
      \rho(\tilde{t})=i
    \right\} \notag \\
    &\qquad\qquad\qquad\qquad\qquad = \pi_{ij}\Delta t+o(\Delta t), \text{
      if $j\ne i$}
    \label{eq:13} \\
    &\Pr\left\{
      P(\rho(\tilde{t}+\Delta t))=P_{i}\mid \xi(\tilde{t}),
      \rho(\tilde{t})=i
    \right\} \notag \\
    &\qquad\qquad\qquad\qquad\qquad = 1+\pi_{ii}\Delta t+o(\Delta t), \text{
      if $j=i$}
    \label{eq:14}
  \end{align}
  In \eqref{eq:11}, $\xi(\tilde{t})$ is considered as the system state just
  before the mode switches. At the switching time, the system state jumps from
  $\xi(\tilde{t})$ to $e^{A_{j}d_{j}} \xi(\tilde{t})$. After the switching,
  the system state change is given by $A_{j}e^{A_{j}d_{j}}
  \xi(\tilde{t})\Delta t + o(\Delta t)$.

  Let $R_{i} \triangleq A_{i}^{T}P_{i}+P_{i}A_{i}+\pi_{ii}P_{i}+ \sum_{j=1,j\ne
    i}^{m}\pi_{ij}e^{A_{j}^{T}d_{j}}P_{j}e^{A_{j}d_{j}}$.  It follows from
  \eqref{eq:11}--\eqref{eq:14} that
  \begin{align*}
    &\quad \E\left\{
      V(\xi(\tilde{t}+\Delta t),\rho(\tilde{t}+\Delta t))
      \mid
      \xi(\tilde{t}), \rho(\tilde{t})=i
    \right\} \\
    &= \E\left\{
      \xi^{T}(\tilde{t}+\Delta t)P(\rho(\tilde{t}+\Delta
      t))\xi(\tilde{t}+\Delta t)
      \mid
      \xi(\tilde{t}), \rho(\tilde{t})=i
    \right\} \\
    &= [1+\pi_{ii}\Delta t]
    [\xi(\tilde{t})+A_{i}\xi(\tilde{t})\Delta t]^{T}
    P_{i}
    [\xi(\tilde{t})+A_{i}\xi(\tilde{t})\Delta t] \\
    &\quad + \sum_{j=1,j\ne i}^{m}\left\{
      [\pi_{ij}\Delta t]
      [\xi(\tilde{t})+A_{j}\xi(\tilde{t})\Delta t]^{T}
      e^{A_{j}^{T}d_{j}}P_{j}e^{A_{j}d_{j}} \right. \\
    &\qquad\qquad\qquad \left.
      \times [\xi(\tilde{t})+A_{j}\xi(\tilde{t})\Delta t]
    \right\}
    + o(\Delta t) \\
    &= \xi^{T}(\tilde{t})[P_{i}+R_{i}\Delta t]\xi(\tilde{t}) 
    + o(\Delta t)
  \end{align*}

  Therefore, the infinitesimal generator of
  $V(\xi(\tilde{t}),\rho(\tilde{t}))$ is given by
  \begin{align}
    &\quad \mathcal{L} V(\xi(\tilde{t}),\rho(\tilde{t})=i) \notag \\
    &\triangleq \lim_{\Delta t\to 0^{+}}
    \frac{1}{\Delta t}
    \left[
      \E
      \left\{
        V(\xi(\tilde{t}+\Delta t),\rho(\tilde{t}+\Delta t))
        \mid \xi(\tilde{t}), \rho(\tilde{t})=i
      \right\}
    \right. \notag \\
    & \qquad\qquad\qquad
    \left.
      - V(\xi(\tilde{t}),\rho(\tilde{t})=i)
    \right]
    \notag \\
    &= \xi^{T}(\tilde{t})R_{i}\xi(\tilde{t})
    = - \xi^{T}(\tilde{t})(-R_{i})\xi(\tilde{t})
    \le -\beta\norm{\xi(\tilde{t})}^{2}
    \label{eq:15}
  \end{align}
  for any $\rho(\tilde{t})=i\in\mathcal{M}$, where $\beta \triangleq
  \min_{i\in\mathcal{M}} \left\{\underline{\sigma}(-R_{i})\right\} >0$.

  From \eqref{eq:15}, we obtain
  \begin{equation*}
    \lim_{\tilde{t}\to\infty} \E
    \left\{
      V(\xi(\tilde{t}),\rho(\tilde{t}))
      \mid
      \xi_{0}, \rho_{0}
    \right\}
    = 0
  \end{equation*}
  and
  \begin{align*}
    \norm{\xi(\tilde{t})}^{2} &\le
    -\frac{1}{\beta} \mathcal{L} V(\xi(\tilde{t}),\rho(\tilde{t}))
  \end{align*}
  for any $\xi(\tilde{t})\ne0$ and any $\rho(\tilde{t})\in\mathcal{M}$. By
  Dynkin's formula~\cite{Oeksendal:2005:book}, we have
  \begin{align*}
    &\quad \E
    \left\{
      \int_{0}^{\infty}\norm{\xi(\tilde{t})}^{2} dt \mid \xi_{0}, \rho_{0}
    \right\} \\
    &\le -\frac{1}{\beta}\E
    \left\{
      \int_{0}^{\infty}\mathcal{L}
      V(\xi(\tilde{t}),\rho(\tilde{t}))dt \mid \xi_{0},\rho_{0}
    \right\} \\
    &= -\frac{1}{\beta}
    \left[
      \E
      \left\{
        \lim_{\tilde{t}\to\infty}
        V(\xi(\tilde{t}),\rho(\tilde{t}))\mid \xi_{0},\rho_{0}
      \right\}
      -V(\xi_{0},\rho_{0})
    \right] \\
    &\le \frac{1}{\beta}V(\xi_{0},\rho_{0}) < \infty
  \end{align*}
  for any $\xi_{0}\ne0$ and $\rho_{0}\in\mathcal{M}$. Therefore, the system
  in~\eqref{eq:5} is stochastically stable, and so is the system
  in~\eqref{eq:1} in view of Lemma~\ref{lem:1}.

  ($\Rightarrow$) 
  According to Lemma~\ref{lem:1}, the stochastic stability of the system
  in~\eqref{eq:1} implies the stochastic stability of the system
  in~\eqref{eq:5}. We will show that there exist matrices
  $P_{i}\in\mathbb{S}^{+}$ such that \eqref{eq:10} holds for all
  $i\in\mathcal{M}$. Let $\xi(\tilde{t})$ be the state trajectory of the
  system in~\eqref{eq:5}. Given any $\xi(\tilde{t}) \ne 0$, $\rho(\tilde{t})
  \in \mathcal{M}$ and $Q_{i} \in \mathbb{S}^{+}$, $i \in \mathcal{M}$, define
  a matrix-valued function $P(\omega-\tilde{t},\rho(\tilde{t}))$ of $\omega$
  and $\rho(\tilde{t})$ such that
  \begin{align}
    \xi^{T}(\tilde{t})P(\omega-\tilde{t},\rho(\tilde{t}))\xi(\tilde{t})
    = 
    \E
    \left\{
      \int_{\tilde{t}}^{\omega}\xi^{T}(\tau)Q(\rho(\tau))\xi(\tau) d\tau
      \mid \xi(\tilde{t}),\rho(\tilde{t})
    \right\}
    \label{eq:16}
  \end{align}
  where $Q(\rho(\tau)=i) \triangleq Q_{i}$. The quadratic form on the left
  side of \eqref{eq:16} is non-decreasing as $\omega$ increases since
  $Q_{i}>0$. It is also bounded from above as $\omega\to\infty$ since the
  system in~\eqref{eq:5} is stochastically stable.  Thus
  $\lim_{\omega\to\infty}\xi^{T}(\tilde{t})P(\omega-\tilde{t},\rho(\tilde{t}))\xi(\tilde{t})$
  exists. We can define a new matrix-valued function $P(\rho(\tilde{t}))$ of
  $\rho(\tilde{t})$ such that
  \begin{align*}
    \xi^{T}(\tilde{t})P(\rho(\tilde{t}))\xi(\tilde{t})
    & =
    \lim_{\omega\to\infty}\xi^{T}(\tilde{t})P(\omega-\tilde{t},\rho(\tilde{t}))\xi(\tilde{t})
    > 0
  \end{align*}
  for any $\xi(\tilde{t})\ne0$ and $\rho(\tilde{t})\in\mathcal{M}$.
  Therefore, we have
  \begin{align}
    P(\rho(\tilde{t}))
    = \lim_{\omega\to\infty}P(\omega-\tilde{t},\rho(\tilde{t}))
    > 0
    \label{eq:17}
  \end{align}
  We have constructed a set of matrices $P_{i} \triangleq P(\rho(\tilde{t})=i) \in
  \mathbb{S}^{+}$ for all $i\in\mathcal{M}$. In the following, we show that
  $P_{i}$ are solutions to \eqref{eq:10}.

  It follows from \eqref{eq:16} that
  \begin{align}
    &\E
    \left\{
      \xi^{T}(\tilde{t})P(\omega-\tilde{t},\rho(\tilde{t}))\xi(\tilde{t})
      - \xi^{T}(\tilde{t}+\Delta t)P(\omega-\tilde{t}-\Delta
      t,\rho(\tilde{t}+\Delta t))
    \right. \notag \\
    &\qquad\quad \left.
      \times \xi(\tilde{t}+\Delta t)
      \mid \xi(\tilde{t}), \rho(\tilde{t})
    \right\} \notag \\
    &=\E
    \left\{
      \E
      \left\{
        \int_{\tilde{t}}^{\omega}\xi^{T}(\tau)Q(\rho(\tau))\xi(\tau) d\tau
        \mid \xi(\tilde{t}), \rho(\tilde{t})
      \right\}
    \right. \notag \\
    &\qquad\qquad
    \left.
      - \E
      \left\{
        \int_{\tilde{t}+\Delta t}^{\omega}\xi^{T}(\tau)Q(\rho(\tau))\xi(\tau)
        d\tau
        \mid \xi(\tilde{t}+\Delta t), \rho(\tilde{t}+\Delta t)
      \right\}
    \right. \notag \\
    &\qquad\qquad\qquad\qquad \left.
      \mid \xi(\tilde{t}), \rho(\tilde{t})
    \right\} \notag \\
    &= \E
    \left\{
      \int_{\tilde{t}}^{\omega}\xi^{T}(\tau)Q(\rho(\tau))\xi(\tau) d\tau
      \mid \xi(\tilde{t}), \rho(\tilde{t})
    \right\} \notag \\
    &\qquad - \E
    \left\{
      \int_{\tilde{t}+\Delta t}^{\omega}\xi^{T}(\tau)Q(\rho(\tau))\xi(\tau) d\tau
      \mid \xi(\tilde{t}), \rho(\tilde{t})
    \right\} \notag \\
    &=\E
    \left\{
      \int_{\tilde{t}}^{\tilde{t}+\Delta t}\xi^{T}(\tau)Q(\rho(\tau))\xi(\tau) d\tau
      \mid \xi(\tilde{t}), \rho(\tilde{t})
    \right\}
    \notag \\
    &= \xi^{T}(\tilde{t})
    \left[
      P(\omega-\tilde{t},i)
      - P(\omega-\tilde{t}-\Delta t,i)
      - A_{i}^{T}P(\omega-\tilde{t}-\Delta t,i)\Delta t
    \right. \notag \\
    &\qquad\quad
    - P(\omega-\tilde{t}-\Delta t,i)A_{i}\Delta t
    - [\pi_{ii}\Delta t]P(\omega-\tilde{t}-\Delta t,i) \notag \\
    &\qquad\quad \left.
      - \sum_{j=1,j\ne i}^{m}
      \left\{
        [\pi_{ij}\Delta t]
        e^{A_{j}^{T}d_{j}}P(\omega-\tilde{t}-\Delta t,j)e^{A_{j}d_{j}}
      \right\}
    \right]\xi(\tilde{t}) \notag \\
    &\qquad + o(\Delta t)
    \label{eq:18} 
  \end{align}
  The second ``$=$'' holds because $(\xi(\tilde{t}),\rho(\tilde{t}))$ is a
  Markov process; the last ``$=$'' follows from the direct calculation of
  \begin{align*}
    &\E
    \left\{
      \xi^{T}(\tilde{t}+\Delta t)P(\omega-\tilde{t}-\Delta t,\rho(\tilde{t}+\Delta t))\xi(\tilde{t}+\Delta t)
      \mid
      \xi(\tilde{t}), \rho(\tilde{t})=i
    \right\} \\
    &= [1+\pi_{ii}\Delta t]
    [\xi(\tilde{t})+A_{i}\xi(\tilde{t})\Delta t]^{T}
    P(\omega-\tilde{t}-\Delta t,i)
    [\xi(\tilde{t})+A_{i}\xi(\tilde{t})\Delta t] \\
    &\quad + \sum_{j=1,j\ne i}^{m}
    \left\{
      [\pi_{ij}\Delta t]
      [\xi(\tilde{t})+A_{j}\xi(\tilde{t})\Delta t]^{T}
      e^{A_{j}^{T}d_{j}}P(\omega-\tilde{t}-\Delta t,j)
    \right. \\
    &\quad\qquad\qquad\quad
    \left.
      \times e^{A_{j}d_{j}}
      [\xi(\tilde{t})+A_{j}\xi(\tilde{t})\Delta t]
    \right\}
    + o(\Delta t)
  \end{align*}

  It follows from \eqref{eq:17} that
  $\lim_{\omega\to\infty}P(\omega-\tilde{t},i)=P_{i}$ and
  $\lim_{\omega\to\infty}P(\omega-\tilde{t}-\Delta t,j)=P_{j}$.  Taking limit
  on the right side of \eqref{eq:18} as $\omega\to\infty$, we have
  \begin{align}
    &\quad \xi^{T}(\tilde{t})
    \left[
      P_{i} - P_{i} - R_{i}\Delta t
    \right]
    \xi(\tilde{t})+o(\Delta t) \notag \\
    &= \E
    \left\{
      \int_{\tilde{t}}^{\tilde{t}+\Delta t}
      \xi^{T}(\tau)Q(\rho(\tau))\xi(\tau) d\tau
      \mid
      \xi(\tilde{t}), \rho(\tilde{t})=i
    \right\}
    \label{eq:20}
  \end{align}
  Dividing both sides of \eqref{eq:20} by $\Delta t$ and taking limit as
  $\Delta t\to0^{+}$, we have
  \begin{align}
    &\quad -\xi^{T}(\tilde{t})
    R_{i}
    \xi(\tilde{t}) \notag \\
    &= \lim_{\Delta t\to0^{+}}\E\left\{
      \frac{\int_{\tilde{t}}^{\tilde{t}+\Delta t}\xi^{T}(\tau)Q(\rho(\tau))\xi(\tau)
        d\tau}{\Delta t}
      \mid
      \xi(\tilde{t}), \rho(\tilde{t})=i
    \right\} \notag \\
    &= \E\left\{
      \lim_{\Delta t\to0^{+}}
      \frac{\int_{\tilde{t}}^{\tilde{t}+\Delta t}\xi^{T}(\tau)Q(\rho(\tau))\xi(\tau)
        d\tau}{\Delta t}
      \mid
      \xi(\tilde{t}), \rho(\tilde{t})=i
    \right\} \notag \\
    &= \E\left\{
      \xi^{T}(\tilde{t})Q(\rho(\tilde{t}))\xi(\tilde{t})
      \mid
      \xi(\tilde{t}), \rho(\tilde{t})=i
    \right\} \notag \\
    &= \xi^{T}(\tilde{t})Q_{i}\xi(\tilde{t})
    \label{eq:21}
  \end{align}
  Because \eqref{eq:21} holds for any $\xi(\tilde{t})$ and
  $\rho(\tilde{t})=i\in\mathcal{M}$, we have $R_{i} = -Q_{i} < 0$.  This
  completes the whole proof.
\end{IEEEproof}

\begin{remark}
  When the fixed dwell time $d_{i}=0$ for all $i\in\mathcal{M}$,
  Theorem~\ref{thm:1} reduces to the well-known stochastic stability result
  for Markovian jump linear systems~\cite{JC:1990:tac}. This is expected as
  the system in~\eqref{eq:5} will reduce to a Markovian jump linear systems if
  $d_{i}=0$ for all $i\in\mathcal{M}$. Actually, the proof of
  Theorem~\ref{thm:1} is inspired by the proof of the stability result for
  Markovian jump systems. Also, the computational complexity of
  Theorem~\ref{thm:1} is the same as the stochastic stability result for
  Markovian jump linear systems.
\end{remark}

\section{Illustrative Example}
\label{sec:illustr-exampl}

In this section, a numerical example is used to illustrate the effect of the
random switching signal on system stability. Some intuitions on system
stability are confirmed by numerical tests. The system data are given in
Table~\ref{tab:exmp:ct}. In Table~\ref{tab:exmp:ct}, $\pi_{12}=\pi_{21}=1$ was
chosen deliberately so that we can focus on the fixed dwell time part. The
stability of these systems was tested for different fixed dwell time values
based on Theorem~\ref{thm:1}.
\begin{table}[hbt]
  \centering
  \caption{System data of three switched systems}
  \label{tab:exmp:ct}
  \begin{tabular}{|c|ccc|}
    \hline
    Case & $A_{1}$ & $A_{2}$ & $\Pi$ \\  \hline
    1 &
    $\Bigl[\begin{array}{cc} -1.2 & 5 \\ 0 & -1 \end{array}\Bigr]$ &
    $\Bigl[\begin{array}{cc} -0.6 &  0 \\ 1 & -0.6 \end{array}\Bigr]$ &
    $\Bigl[\begin{array}{cc} -1 & 1 \\ 1 & -1 \end{array}\Bigr]$ \\ \hline
    2 &
    $\Bigl[\begin{array}{cc} -1 & 0 \\ 1 & -1  \end{array}\Bigr]$ &
    $\Bigl[\begin{array}{cc} 0.3 & 0.1 \\ 0 & 0.2 \end{array}\Bigr]$ &
    $\Bigl[\begin{array}{cc} -1 & 1 \\ 1 & -1 \end{array}\Bigr]$ \\ \hline
    3 &
    $\Bigl[\begin{array}{cc} -0.5 & 0 \\ 0.1 & 0.4  \end{array}\Bigr]$ &
    $\Bigl[\begin{array}{cc}  0.3 & 1.5 \\ 0 & -3  \end{array}\Bigr]$ &
    $\Bigl[\begin{array}{cc} -1 & 1 \\ 1 & -1 \end{array}\Bigr]$ \\ \hline
  \end{tabular}
\end{table}

The stability results are depicted in Fig.~\ref{fig:e21}, Fig.~\ref{fig:e22}
and Fig.~\ref{fig:e23}, respectively, where the shaded areas are the stability
regions.  In Case~1, the subsystems are chosen to be stable, the stability
test result in Fig.~\ref{fig:e21} shows that the stability region is a
non-convex set and that increasing the fixed dwell time (either $d_{1}$ or
$d_{2}$ or both) will stabilize the system. This result suggests that slow
switching is recommended when all the subsystems are stable. Case~2
corresponds to the situation where one subsystem is stable and the other is
unstable; here $A_{1}$ is stable and $A_{2}$ unstable; the stability test
result in Fig.~\ref{fig:e22} shows that the stable and unstable regions divide
the first orthant into two semi-infinite parts, and the boundary between the
two regions looks like a straight line. The result also confirms our
expectation: increasing the dwell time in the stable subsystem will stabilize
the system and increasing the dwell time in the unstable subsystem will
destabilize it.  In Case~3, the two subsystems are unstable. The test result
in Fig.~\ref{fig:e23} shows that the stability region looks like a closed
convex set and that both slow and fast switching will destabilize the
system. Therefore, the dwell time needs to be chosen carefully to make the
system stable in this case.

Finally, there are totally six variables in the computations of this example.
The computational complexity is the same as the stability test for Markovian
jump linear systems.

\section{Conclusions}
\label{sec:conclusions}

This paper studied the stability property of randomly switched systems where
the dwell time in each subsystem consists of a fixed part and a random
part. We first showed the stochastic stability of such systems is equivalent
to the stochastic stability of a class of Markovian jump systems with state
jumps at the mode switching times.  Then a necessary and sufficient condition
for the system stability was derived using a stochastic Lyapunov
approach. Finally, a numerical example was used to illustrate the application
of the theory, and the results are consistent with our intuitions. Future
research could be directed to the development of numerical algorithms to find
the stabilizing switching parameters for the switched systems, and to the
stability analysis of randomly switched singular systems, randomly switched 2D
systems, randomly switched time-delay systems.

\appendix

\label{sec:two-lemmas}

We provide two lemmas that are used in the proof of Lemma~\ref{lem:1}.

\begin{lemma}
  \label{lem:2}
  Given a matrix $A\in\mathbb{R}^{n\times n}$, any vector
  $x\in\mathbb{R}^{n}$. Then there exists a constant $c_{0}\in\mathbb{R}$,
  which is independent of $x$, such that $\norm{e^{At}x}^{2}\ge
  e^{c_{0}t}\norm{x}^{2}$ for any $t\in[0,\infty)$. Moreover, given any
  $c\in(-\infty,c_{0}]$, $\norm{e^{At}x}^{2}\ge e^{ct}\norm{x}^{2}$ for
  $t\in[0,\infty)$.
\end{lemma}
\begin{IEEEproof}
  Let $c_{0}=\lambda_{\mathrm{min}}(A+A^{T})$. Then
  $A+A^{T}-c_{0}I\ge0$. Consider the function
  $g(t)\triangleq\norm{e^{(A-\frac{c_{0}}{2}I)t}x}^{2}$, we have
  \begin{align*}
    \frac{dg(t)}{dt}
    &= \frac{d}{dt}
    \left\{
      x^{T}e^{(A-\frac{c_{0}}{2}I)^{T}t}e^{(A-\frac{c_{0}}{2}I)t}x
    \right\} \\
    &=
    x^{T}e^{(A^{T}-\frac{c_{0}}{2}I)t}
    \left[
      A+A^{T}-c_{0}I
    \right]
    e^{(A-\frac{c_{0}}{2}I)t}x \\
    &\ge 0
  \end{align*}
  Hence $g(t)\ge g(0)$ for $t\in[0,\infty)$; that is,
  $\norm{e^{(A-\frac{c_{0}}{2}I)t}x}^{2}\ge\norm{x}^{2}$. Therefore, $
  \norm{e^{At}x}^{2} = \norm{e^{(A-\frac{c_{0}}{2}I)t} e^{\frac{c_{0}}{2}It}
    x}^{2} \ge e^{c_{0}t}\norm{x}^{2}\ge e^{ct}\norm{x}^{2} $ for any
  $t\in[0,\infty)$ and any $c\le c_{0}$.
\end{IEEEproof}

\begin{lemma}
  \label{lem:3}
  Given an exponentially distributed random variable $X$ with parameter
  $\lambda$, and two real numbers $a<\lambda$ and $b\in\mathbb{R}$. Then
  $\E\left\{\int_{b}^{b+X}e^{at}dt\right\}=\frac{e^{\lambda b}}{\lambda-a}$.
\end{lemma}

\begin{IEEEproof}
  The result follows from direct computation. That is,
  \begin{align*}
    &\quad \E\left\{\int_{b}^{b+X}e^{at}dt\right\} \triangleq
    \int_{0}^{\infty} \left[\int_{b}^{b+x}e^{at}dt\right] \lambda e^{-\lambda x} dx \\
    &= \int_{b}^{\infty} \left[\int_{t-b}^{\infty} \lambda e^{-\lambda x}dx
    \right] e^{at} dt 
    = \int_{b}^{\infty} e^{-\lambda(t-b)} e^{at} dt \\
    &= e^{\lambda b}\int_{0}^{\infty} e^{(a-\lambda)t} dt 
    = \frac{e^{\lambda b}}{\lambda-a} 
  \end{align*}
\end{IEEEproof}



\begin{figure}[hbt]
  \centering
  \includegraphics[width=7cm]{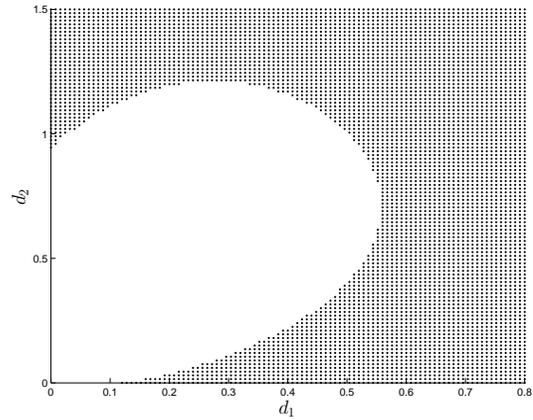}
  \caption{Case 1: stability region of a switched system composed of two
    stable subsystems. The shaded area is the stability region. It shows that
    slow switching is recommended in this case.}
  \label{fig:e21}
\end{figure}

\begin{figure}[hbt]
  \centering
  \includegraphics[width=7cm]{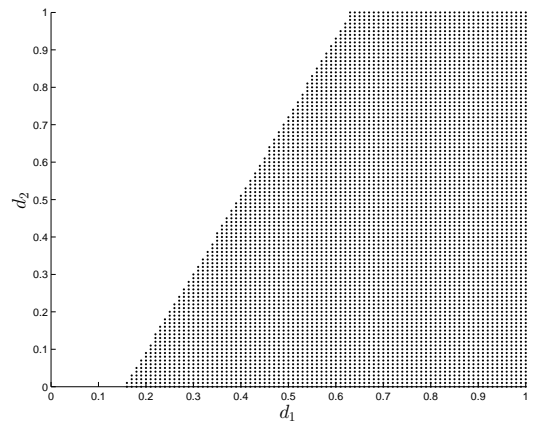}
  \caption{Case 2: stability region of a switched system composed of a stable
    subsystem and an unstable subsystem. The shaded area is the stability
    region. It shows that increasing the dwell time in the stable subsystem
    will stabilize the system.}
  \label{fig:e22}
\end{figure}

\begin{figure}[hbt]
  \centering
  \includegraphics[width=7cm]{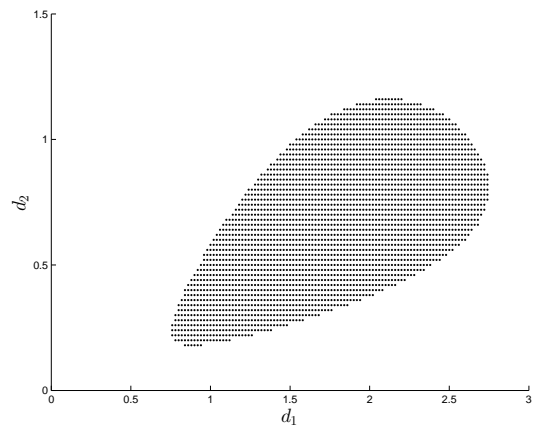}
  \caption{Case 3: stability region of a switched system composed of two
    unstable systems. The shaded area is the stability region. It shows that
    either slow or fast switching can destabilize the system and that a
    careful chosen dwell times can stabilize the system.}
  \label{fig:e23}
\end{figure}

\end{document}